**Title**

# NeuroInspector: A Local-First Environment for Inspecting and Annotating Hierarchical Neuroscience Datasets

**Author**

Zihan Yang[1]*

**Affiliations**

[1]Department of Biosciences, Durham University, Durham, United Kingdom

*Correspondence: zihan.yang@durham.ac.uk

## Abstract

The growing scale and structural complexity of neuroscience datasets have made dataset inspection an increasingly distinct stage of the research workflow. Existing inspection workflows, however, remain fragmented, often relying on exploratory scripts, manual documentation, and repeated navigation of unfamiliar file structures before meaningful scientific analysis can begin. Here we present NeuroInspector, a lightweight, browser-based environment for inspecting HDF5 and NWB files. Using WebAssembly-based HDF5 parsing (h5wasm), NeuroInspector runs entirely client-side: files are opened directly from local disk, and the application implements no file-upload endpoint, including in its hosted deployment. The tool combines structural navigation, metadata inspection, sampled data previews, and path-level annotation into portable, fingerprinted "project packs" that preserve inspection decisions without modifying the original file. Rather than functioning as an analysis or validation platform, NeuroInspector provides a dedicated, traceable environment for the inspection stage that precedes formal analysis.

## Introduction

Modern neuroscience increasingly relies on large, multimodal datasets generated from diverse experimental modalities, including calcium imaging, electrophysiology, behavioral monitoring, and anatomical profiling. HDF5 provides a flexible hierarchical storage format for organizing these heterogeneous data [1], while Neurodata Without Borders (NWB) [2] builds on the Hierarchical Data Modeling Framework (HDMF) [3] to standardize the organization of neurophysiological data and metadata, typically stored in HDF5 as its container format. Together, these formats facilitate data sharing, long-term preservation, and reproducible analyses in the spirit of the FAIR principles [4]. As datasets continue to grow in both size and structural complexity, understanding their internal organization has become an essential prerequisite for downstream scientific analysis.

Before meaningful analysis can begin, researchers must first familiarize themselves with an unfamiliar dataset. This process typically involves identifying relevant data groups, locating variables of interest, interpreting metadata, examining dataset attributes, and establishing the relationships between stored objects and experimental variables. Although this inspection stage rarely contributes directly to scientific discovery, it is indispensable for every subsequent analytical step. In practice, it is often performed through a combination of exploratory scripts, repeated command-line queries, manual note-taking, and iterative navigation of hierarchical file structures. As neuroscience datasets become increasingly complex, this preliminary inspection has evolved from a trivial preparatory task into an independent stage of the research workflow.

A number of existing tools already make HDF5 or NWB data browsable — desktop viewers such as HDFView [5], browser-based tools such as h5web [6] and myHDF5 [7], together with h5web's integrations for Visual Studio Code and JupyterLab (vscode-h5web, jupyterlab-h5web) [8], and command-line validators such as NWB Inspector [9]. NeuroInspector is not meant to replace these; it addresses a narrower, complementary need — a fully client-side environment for inspecting HDF5 files and HDF5-backed NWB files that pairs structural inspection with a portable, source-fingerprinted record of what was found and why it mattered. To our knowledge, these tools do not combine path-level annotation with portable, source-fingerprinted inspection records in a standalone, zero-install browser workflow.

Here, we present NeuroInspector, a lightweight, local-first environment designed to streamline the inspection of hierarchical neuroscience datasets. NeuroInspector enables interactive exploration of HDF5 and NWB files directly within a web browser without software installation or remote data upload. Beyond visualizing hierarchical structures, it supports metadata inspection, sampled dataset preview, path-level annotation, and the creation of portable project records that preserve selected paths and inspection notes for subsequent analyses. Rather than introducing another data analysis platform, NeuroInspector focuses on reducing the friction between receiving an unfamiliar dataset and beginning scientific analysis.

### Design Principles

NeuroInspector was designed around the practical requirements of neuroscience data inspection rather than general-purpose file management or data analysis. **Figure 1** contrasts the conventional, fragmented inspection process — scattered across temporary scripts, notebooks, file explorers, screenshots, and manual notes, with important paths and observations easily lost and researchers repeatedly returning to the source file — with the continuous, six-step workflow NeuroInspector is built to support, and summarizes the four design principles behind it: local-first execution, zero-install accessibility, structure-centered inspection, and traceable inspection.

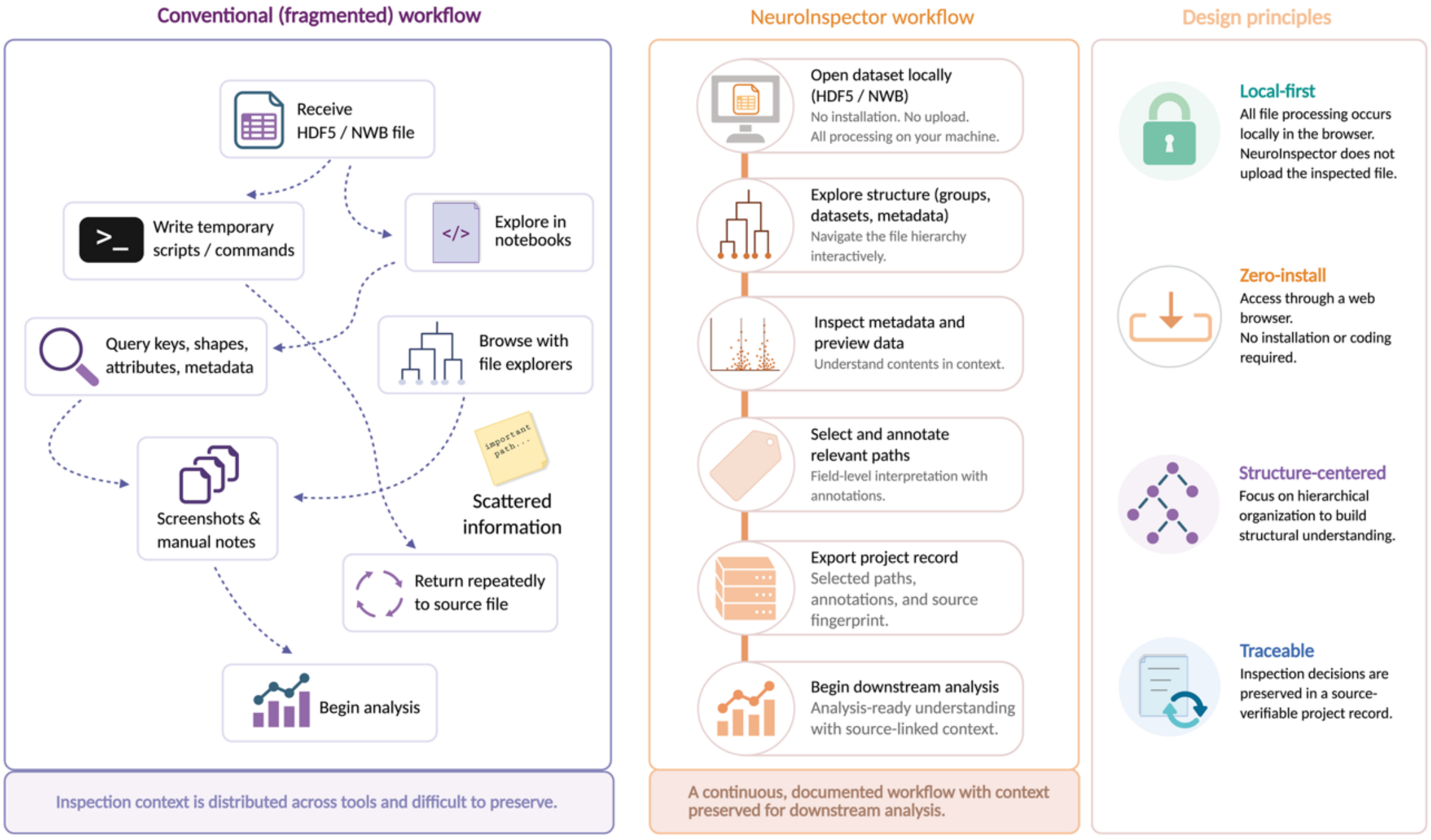

**Figure 1. From a fragmented inspection process to a continuous, traceable workflow. Left**, conventional inspection is distributed across temporary scripts, notebooks, viewers, screenshots, and manual notes, making

relevant paths and interpretations difficult to preserve. **Center**, NeuroInspector organizes local file opening, structural exploration, metadata and value inspection, path annotation, project-record export, and transition to downstream analysis into a continuous workflow. **Right**, the workflow follows four design principles: local-first execution, zero-install accessibility, structure-centered inspection, and source-verifiable documentation.

***Local-first execution.*** Neuroscience datasets are frequently large, institutionally managed, or subject to data-sharing restrictions, making remote upload impractical or undesirable in many research environments. NeuroInspector adopts a local-first design in which all file parsing, visualization, annotation, and project management are performed entirely on the user's local machine via WebAssembly: the application implements no file-upload endpoint, and inspected files are processed locally in the browser. This design improves privacy and avoids network transfer of the inspected dataset, making the software suitable for routine use with sensitive or proprietary datasets.

***Zero-install accessibility.*** The first encounter with an unfamiliar dataset should require as little setup as possible. NeuroInspector is deployed as a hosted web application: researchers can begin inspecting a local file immediately, without installing software, configuring an environment, or writing a line of code.

***Structure-centered inspection.*** Rather than emphasizing numerical visualization alone, NeuroInspector is designed around the hierarchical structure of HDF5 and NWB datasets. Interactive navigation across an overview panel, a full file-tree view, and a fast quick-tree view allows researchers to progressively reconstruct the semantic organization of unfamiliar files. Dataset previews, metadata inspection, and structural navigation are complementary components supporting the same objective: transforming complex hierarchical storage into an interpretable representation of experimental organization.

***Traceable inspection.*** During inspection, researchers identify variables of interest, interpret metadata, select relevant datasets, and establish relationships between experimental concepts and file structures. NeuroInspector allows selected paths and free-text notes to be preserved as a portable, source-verifiable project pack linked to the original dataset via a content fingerprint, turning exploratory observations into a persistent, revisitable record.

## Implementation

NeuroInspector is implemented as a single-page web application (React 19, TypeScript, Vite) that runs entirely in the browser. HDF5 files and HDF5-backed NWB files are parsed client-side using h5wasm [10], a WebAssembly build of the HDF5 C library. NeuroInspector implements no file-upload endpoint: the selected HDF5 or HDF5-backed NWB file is processed locally in the browser, and the application's hosted GitHub Pages deployment serves only static assets. The hosted application is designed for modern desktop browsers with WebAssembly support, including Chrome, Edge, Firefox, and Safari, and needs no local installation; building it from source instead requires Node.js 22+.

Structural navigation is not sampled: the application traverses the full available hierarchy of groups and datasets and reports each entry's structural fields — shape, dtype, chunking, compression — and attributes. Long textual values may be visually truncated, with complete content available on hover; this display behavior does not affect structural traversal. Sampling proper applies only to the numeric preview of array values, which can be arbitrarily large: because full arrays can exceed what a browser can safely render, NeuroInspector displays bounded, sampled previews of values rather than the complete array. Overview and grouped previews are capped at approximately 32,000 elements, and the dedicated dataset view supports up to approximately 250,000 elements; two-dimensional arrays are rendered as tables when small (≤24×24, ≤256 cells) and as heatmaps otherwise. These budgets keep the interface responsive but mean that previews are sampled approximations and should not be treated as a substitute for reading the full array in an analysis pipeline — a point the interface flags directly to the user.

Inspection outcomes are preserved as a project pack: a portable JSON file (.neuroinspector.json, schema neuroinspector-project/v1) recording the SHA-256 fingerprint of the source file, the set of selected dataset/group paths, and any annotations, without altering the original HDF5/NWB bytes. Re-opening the same file lets the application verify, via the fingerprint, whether a previously exported project pack still corresponds to its source; a mismatch is surfaced to the user rather than silently accepted. File-tree hierarchies can additionally be exported independently as portable JSON or a self-contained, expandable HTML snapshot.

The application state is managed with Zustand, and visualizations (series, distributions, heatmaps) are rendered with Recharts and a canvas-based heatmap component.

**Inspection Workflow**

The NeuroInspector workflow in **Figure 1** breaks this process into six steps, from opening a file to beginning downstream analysis.

***Open dataset locally.*** Inspection begins when researchers open a local HDF5 or NWB file through the browser's file picker or by drag-and-drop. No installation or upload precedes this step; all processing happens on the user's machine.

***Explore structure.*** The Overview panel surfaces file-level summaries (size, group/dataset counts, root attributes, largest dataset), while the File tree and Quick tree give interactive navigation of the file hierarchy, with structural metadata and attributes exposed for each entry inspected.

***Inspect metadata and preview data.*** Selecting an entry shows its structural metadata — shape, dtype, chunking, compression, and attributes — alongside a sampled preview of its values, letting researchers understand what a dataset contains before committing to it as an analysis target.

***Select and annotate relevant paths.*** Researchers mark relevant groups and datasets and attach free-text notes to each, capturing the reasoning behind a selection rather than the selection alone.

***Export project record.*** Selected paths and annotations are collected into a project pack: a portable, fingerprinted JSON file (.neuroinspector.json) that travels independently of the original dataset.

***Begin downstream analysis.*** Researchers move into their existing tools — PyNWB, HDMF, MATLAB, R, or others — with a structured understanding of the file and a project pack that can be reopened later, reconnected to the source file via its fingerprint, and shared with collaborators.

**Workflow Demonstration**

All demonstrations reported here were performed using NeuroInspector v0.1.0. NeuroInspector was applied to a publicly available intracellular (patch-clamp) electrophysiology NWB file, sub-710327122_ses-711201882_icephys.nwb, drawn from the DANDI Archive. Opening the 16.8 MB file (**Figure 2A**) produced an overview of 349 groups, 680 datasets, and five root attributes (**Figure 2B**). The focused acquisition entry contained 1,701,000 little-endian floating-point values with an approximately 2 MB compressed storage footprint (shuffle-and-deflate compression); NeuroInspector reported its structural metadata without sampling and rendered a bounded preview of the first 32,000 values as a time series and a value distribution, consistent with the sampling budget described in Implementation.

Detailed inspection in the File tree (**Figure 2C**) exposed attributes specific to the source acquisition software (IGORWaveNote, IGORWaveScaling, IGORWaveType) alongside standard fields such as conversion, resolution, and unit. At this point in the session, four of the six entries under a single acquisition sweep — capacitance_fast, capacitance_slow, data, and electrode — had been marked for inclusion in a running project summary.

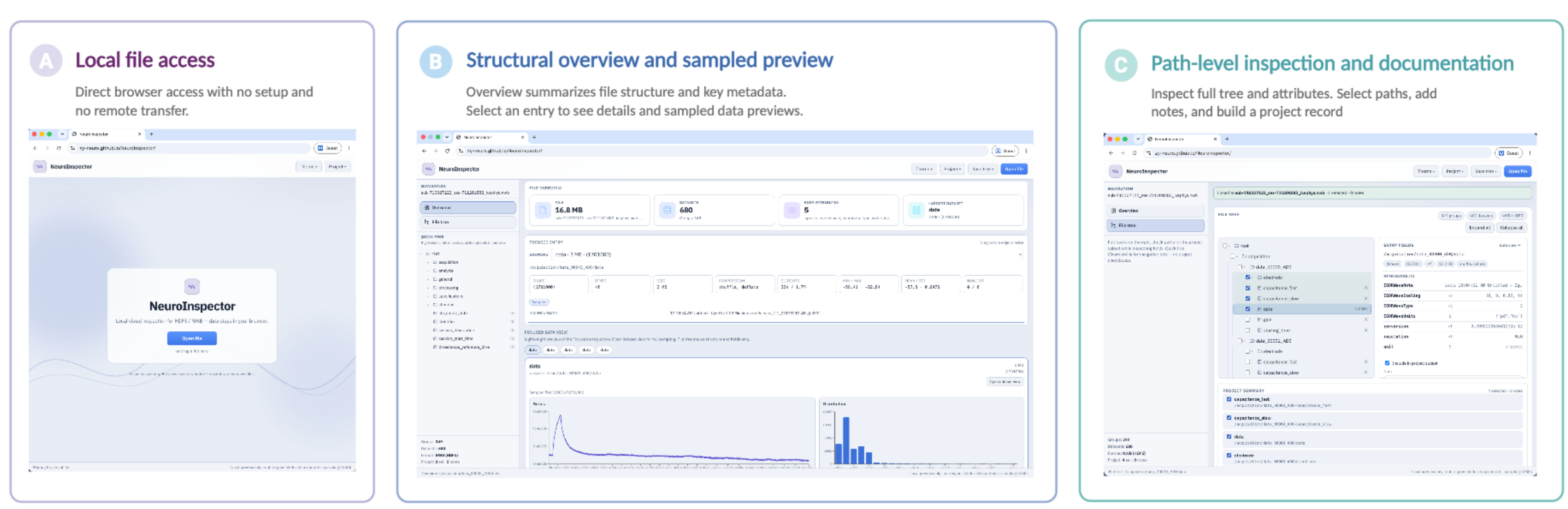

**Figure 2. The NeuroInspector interface applied to a public intracellular electrophysiology dataset. (A)** Local file access: the application opens directly in the browser to a landing page with no setup and no remote data transfer, and a local file is opened via a button or by drag-and-drop. **(B)** Structural overview and sampled preview: after opening the file, the Overview panel reports file-level summary statistics (16.8 MB, 349 groups, 680 datasets, 5 root attributes) and, for the currently focused entry, its structural metadata alongside a sampled preview — here the first 32,000 of 1,701,000 floating-point samples of an acquisition trace (approximately 2 MB compressed on disk), shown as both a time series and a value distribution. **(C)** Path-level inspection and documentation: the File tree exposes the complete hierarchy and entry-level attributes,

including acquisition-software-specific fields (e.g., IGORWaveNote, IGORWaveScaling, IGORWaveType) alongside standard fields such as conversion, resolution, and unit; researchers mark relevant paths for inclusion in a project record, shown here with four of six related entries — capacitance_fast, capacitance_slow, data, and electrode — selected within a single acquisition sweep.

By the time the inspection record was exported (**Figure 3**), the selection had been extended to the full acquisition sweep — all six child entries (electrode, capacitance_fast, capacitance_slow, data, gain, starting_time) together with their parent group, seven paths in total — and collected into a project pack: a .neuroinspector.json file recording the format version, an export timestamp, the source file's identity (name, size in bytes, and a SHA-256 fingerprint), the seven selected paths, and an annotations array, empty in this particular export since no free-text notes had been added. Re-opening the source file lets the recorded fingerprint be checked against the file's current bytes; on a match, the selections are restored and inspection can continue. The same JSON file can also be read directly by external tools — Python, MATLAB, or R — to carry the selected paths and any associated annotations into a separate analysis pipeline.

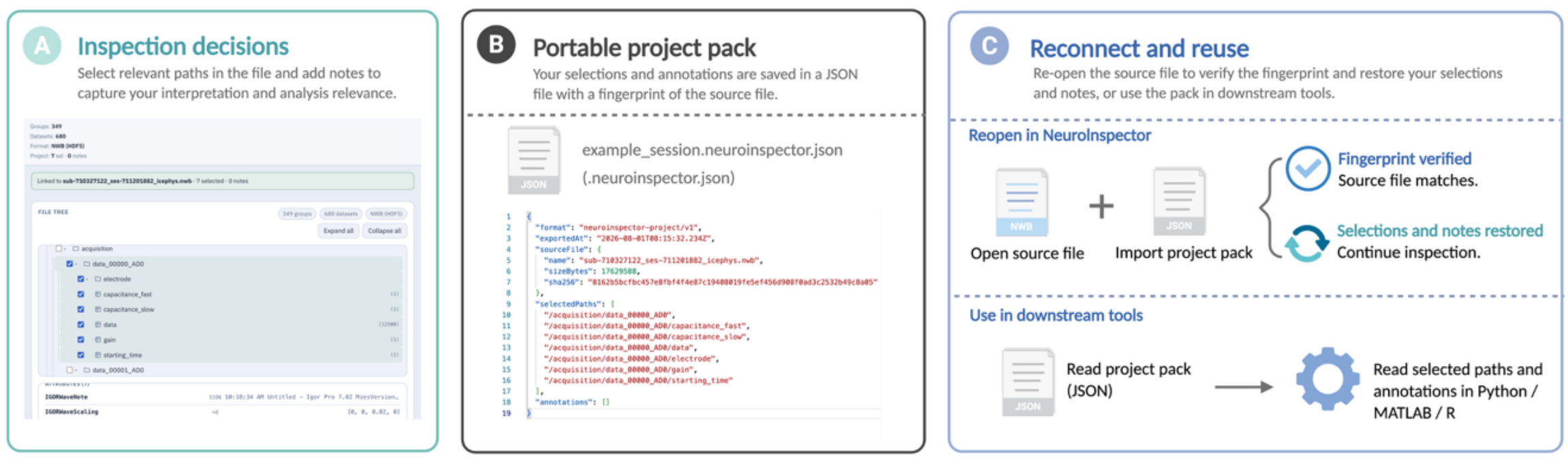

**Figure 3. The portable project pack, from selection to reuse. (A)** Inspection decisions: relevant paths are selected directly in the File tree, here seven paths spanning one acquisition sweep and its six child entries (electrode, capacitance_fast, capacitance_slow, data, gain, starting_time), with a free-text note available on each. **(B)** Portable project pack: selected paths and any associated annotations are exported as a single JSON file (.neuroinspector.json) recording the format version, an export timestamp, the source file's name, size, and SHA-256 fingerprint, the list of selected paths, and an annotations array (empty in this example, since no notes were added for this export). **(C)** Reconnect and reuse: re-opening the source file and importing the project pack lets NeuroInspector verify the fingerprint and restore the recorded selections, or the JSON file can be read directly by external tools — Python, MATLAB, or R — to carry the selected paths and any associated annotations into a separate analysis pipeline.

## Discussion

The growing scale and structural complexity of hierarchical neuroscience datasets has introduced a research stage that is often overlooked but increasingly essential: understanding the organization of unfamiliar data before formal analysis begins. NeuroInspector was developed to support this stage by combining local, browser-based HDF5/NWB parsing with structural navigation, sampled preview, and traceable annotation, rather than treating inspection as a collection of temporary exploratory tasks.

An important aspect of this workflow is that it emphasizes structural understanding rather than numerical analysis. By enabling researchers to navigate, interpret, annotate, and document hierarchical datasets within a single environment, NeuroInspector shifts effort from understanding file organization toward understanding the scientific content the files represent. In this sense, the software is intended to complement, rather than replace, existing viewers such as h5web/myHDF5 and validators such as NWB Inspector, and downstream tools such as PyNWB, HDMF, MATLAB, and R-based analysis pipelines.

A natural question is why inspection outcomes need a dedicated project-pack format rather than, say, a plain-text file of notes. Free-form notes alone do not provide a machine-verifiable link to the file they describe, particularly when files are renamed, relocated, duplicated, or reprocessed. Because the fingerprint is derived from file content rather than file name or location, the relationship remains verifiable after the source file is renamed or relocated. The project pack can then be reopened in NeuroInspector to restore the recorded path selections and annotations, or parsed programmatically by downstream scripts. It is a deliberate trade-off: the structured schema constrains how inspection outcomes are represented, but makes them portable, verifiable, and machine-readable, in keeping with broader calls for reproducible computational practice [11].

***Limitations.*** The current release (v0.1.0) is intentionally narrow: NeuroInspector supports inspection and annotation but does not validate, analyze, or modify the original file. File size is constrained by the memory available to the browser. Planned capabilities, including executable subset export and NWB best-practice validation, are not part of the current release.

As neuroscience datasets continue to increase in scale and structural complexity, improving the accessibility and reproducibility of dataset inspection may become as important as improving the analytical methods applied to the data themselves.

## Software Availability

NeuroInspector is open-source software released under the **Apache License, Version 2.0**.

- **Source code**: https://github.com/zy-neuro/NeuroInspector
- **Live application (no installation required)**: https://zy-neuro.github.io/NeuroInspector/
- **Archived release**: Zenodo. https://doi.org/10.5281/zenodo.21729447 [12]
- **Requirements**: A modern desktop browser with WebAssembly support (Chrome, Edge, Firefox, or Safari). Building from source additionally requires Node.js 22 or later (optionally via the provided Conda environment specifying Python 3.12 and Node.js 22).
- **Supported formats**: .h5, .hdf5, and .nwb (current release: v0.1.0).

## Data Availability

No new experimental data are reported in this manuscript, and no new human or animal data were collected for this study. The demonstration dataset shown in Figures 2–3 (sub-710327122_ses-711201882_icephys.nwb, an intracellular patch-clamp electrophysiology recording) is drawn from Dandiset 000020, *Patch-seq recordings from mouse visual cortex* (Allen Institute for Brain Science, version 0.210913.1639), **DANDI Archive**, https://doi.org/10.48324/dandi.000020/0.210913.1639 [13]

## Acknowledgements

The author thanks the open-source communities behind h5wasm, React, and the broader HDF5/NWB ecosystem that made this work possible.

## Author Contributions

Z.Y. conceived the project, designed and implemented the software, performed the demonstration, prepared the figures, and wrote the manuscript.

**Competing Interests**

The author declares no competing interests.